\documentclass[aps,pra,twocolumn,amsmath,amssymb,nofootinbib,superscriptaddress]{revtex4}

\newcommand{\bra}[1]{\langle#1|}
\newcommand{\ket}[1]{|#1\rangle}

\usepackage[pdftex]{graphicx}
\usepackage{mathrsfs}

\begin{document}

\bibliographystyle{apsrev}

\title{Increasing the dimensionality of quantum walks using multiple walkers}

\author{Peter P. Rohde}
\email[]{dr.rohde@gmail.com}
\homepage{http://www.peterrohde.org}
\affiliation{University of Paderborn, Applied Physics, 33098 Paderborn, Germany}
\affiliation{Centre for Engineered Quantum Systems, Department of Physics and Astronomy, Macquarie University, Sydney NSW 2113, Australia}

\author{Andreas Schreiber}
\affiliation{University of Paderborn, Applied Physics, 33098 Paderborn, Germany}
\affiliation{Max Planck Institute for the Science of Light, 91058 Erlangen, Germany}

\author{Martin {\v S}tefa{\v n}{\' a}k}
\affiliation{Department of Physics, Faculty of Nuclear Sciences and Physical Engineering, Czech Technical University in Prague, Czech Republic}

\author{Igor Jex}
\affiliation{Department of Physics, Faculty of Nuclear Sciences and Physical Engineering, Czech Technical University in Prague, Czech Republic}

\author{Alexei Gilchrist}
\affiliation{Centre for Engineered Quantum Systems, Department of Physics and Astronomy, Macquarie University, Sydney NSW 2113, Australia}

\author{Christine Silberhorn}
\affiliation{University of Paderborn, Applied Physics, 33098 Paderborn, Germany}
\affiliation{Max Planck Institute for the Science of Light, 91058 Erlangen, Germany}

\date{\today}

\frenchspacing

%
%

\begin{abstract}
We show that with the addition of multiple walkers, quantum walks on a line can be transformed into lattice graphs of higher dimension. Thus, multi-walker walks can simulate single-walker walks on higher dimensional graphs and vice versa. This exponential complexity opens up new applications for present-day quantum walk experiments. We discuss the applications of such higher-dimensional structures and how they relate to linear optics quantum computing. In particular we show that multi-walker quantum walks are equivalent to the {\sc BosonSampling} model for linear optics quantum computation proposed by Aaronson \& Arkhipov. With the addition of control over phase-defects in the lattice, which can be simulated with entangling gates, asymmetric lattice structures can be constructed which are universal for quantum computation.
\end{abstract}

\maketitle

%
%

\section{Introduction}

Quantum walks \cite{bib:ADZ, bib:AAKV, bib:Kempe08, bib:Salvador12} (QWs), the quantum analogue of classical random walks, have emerged as an interesting approach to implementing quantum information processing tasks, and provide a useful tool for algorithm design \cite{bib:Childs09}. Recently there have been numerous experimental demonstrations of optical QWs \cite{bib:Hagai08, bib:Schreiber10, bib:Broome10, bib:Peruzzo10, bib:Schreiber11b, bib:Matthews11, bib:Schreiber12, Sansoni12}. Schreiber \emph{et al.} \cite{bib:Schreiber10, bib:Schreiber11b} demonstrated a highly scalable optical QW on a line, whereby a single walker, simulated by weak coherent light, was temporally encoded and the size of the walk was limited only by loss. Shortly after, Broome \emph{et al.} \cite{bib:Broome10} demonstrated a single photon QW  comprising an interferometer constructed from wave-plates and birefringent crystals. A fully integrated device was demonstrated by Peruzzo \emph{et al.} \cite{bib:Peruzzo10} in an interferometrically stable wave-guide. The state of the art is an experiment by Schreiber \emph{et al.} \cite{bib:Schreiber12}, who demonstrated a single walker quantum walk on a 2D lattice using temporally encoded position states.

A central question in such demonstrations is how to scale the system to achieve the exponential complexity required for many interesting quantum information processing (QIP) \cite{bib:NielsenChuang00} applications. This paper examines the ability to exponentially increase the complexity of such systems with the introduction of multiple walkers. The work by Peruzzo \emph{et al.} \cite{bib:Peruzzo10} was a first step in experimentally achieving such complexity, and subsequently Refs. \cite{bib:Matthews11,bib:Owens11}. However the pressing question is to understand how this can be employed for QIP tasks \cite{bib:Guzik, bib:Berry11, Sansoni12}. Understanding such complexity is the central question we will address in this paper.

Recently Rohde \emph{et al.} \cite{bib:RohdeSchreiber11} presented a formalism for multi-walker QWs. In this paper we extend this formalism and examine the applications for such higher-dimensional walks. In particular, we demonstrate that the introduction of multiple walkers is equivalent to employing exponentially larger graph structures, one of the key ingredients in implementing QWs that cannot be efficiently classically simulated, thereby opening the field to experiments of genuine interest to QIP. We demonstrate that multi-walker QWs are useful in simulating higher dimensional single-walker walks. Conversely, a single-walker walk on a higher dimensional graph structure presents us with a versatile approach to mimicking multi-walker walks on simpler graphs. Such a simulation presents us with the tools necessary to understand multi-walker interactions, which are otherwise experimentally challenging to implement. This is particularly attractive to optical implementations, where directly implementing multi-walker interactions is difficult, requiring massive optical non-linearities or entangling gates \cite{bib:KLM01}. In light of the experiments by Peruzzo \emph{et al.} \cite{bib:Peruzzo10} and Matthews \emph{et al.} \cite{bib:Matthews11} we also give consideration to entangled input states and discuss how they may, or may not, benefit experiments.

There are two general classes of quantum walks that have been shown to be universal for quantum computation: continuous- (by Childs \cite{bib:Childs09}) and discrete-time (by Lovett \emph{et al.} \cite{bib:Lovett10}) QWs. In this paper we will focus on discrete-time QWs. We demonstrate that multiple walkers generate a \emph{virtual graph}, whose dimensionality is higher than that of the underlying graph. For example, two walkers transforms a linear graph into a regular lattice in two dimensions, and with three walkers into a three-dimensional lattice. Thus, as the number of walkers increases, the size of the virtual graph increases exponentially. This is an interesting observation, as it implies that with frugal physical resources we can construct a system of exponential complexity.

In particular, we demonstrate that a multi-walker QW is isomorphic to the recently introduced {\sc BosonSampling} model for optical quantum computation (QC) introduced by Aaronson \& Arkhipov \cite{bib:AaronsonArkhipov10}. This model is strongly believed to not be universal for QC, but is nonetheless believed to be classically hard to simulate, making it of direct interest to optical QC experimentalists. However, when lattice defects are introduced, which can be simulated with entangling gates, the scheme becomes universal. Furthermore, we argue that a multi-walker QW on an efficiently sized graph (equivalently {\sc BosonSampling}) is universal for QC only if a QW on an exponentially large graph with $N$ vertices is universal with control over just $O(\log\, N)$ coins, which would be a surprising result, giving further weight to the argument that {\sc BosonSampling} is not universal for QC. However, efficient QC is possible with the addition of CPHASE gates, which will be discussed. We also discuss the potential to using entangled state preparation to overcome the difficulty of implementing entangling operations within a QW.

%
%

\section{The quantum walk formalism}

A QW consists of a \emph{walker}, which may occupy some number of \emph{position} states in a graph. In a discrete-time QW, a walker is a bipartite system consisting of a \emph{position} ($x$) and a \emph{coin} ($c$) value, $\ket{x,c}$. The position literally corresponds to the walker's position in the graph, i.e. at which vertex the walker resides, whereas the coin is an ancillary parameter whose value determines the direction of propagation of the walker. The walk proceeds using two operations -- \emph{coin} ($C$) and \emph{step} ($S$). $C$ coherently randomises the value of the coin parameter, whereas $S$ uses the value of the coin to propagate the walker. In a classical random walk the evolution proceeds by flipping a classical coin and then depending on its value propagating the walker in the appropriate direction. In a QW, on the other hand, the coin may be in a coherent superposition. Thus after propagation the walker is, in general, in a superposition of different positions.

As an example, we consider the archetypal example of a quantum walk -- a single-walker on a linear graph. Then, using a so-called Hadamard coin, the evolution proceeds as,
\begin{eqnarray}
C\ket{x,\pm 1} &=& (\ket{x,1}\pm\ket{x,-1})/\sqrt{2}, \nonumber \\
S\ket{x,c} &=& \ket{x+c,c}.
\end{eqnarray}
Evidently, at each step, the walker enters a superposition of the position states immediately to its left and right. The QW has some interesting features compared to classical walks, including localisation and an enhanced rate-of-spread of the walker \cite{bib:Kempe08}. The increased rate of spread of a QW compared to the classical counterpart is one of the features of QWs that makes it of direct interest to certain algorithmic applications such a search algorithms.

%
%

\section{The multi-walker quantum walk formalism}

Following Ref. \cite{bib:RohdeSchreiber11}, we represent multiple walkers on arbitrary graph structures using \emph{walker operators}. Formally, we define a walker operator of the form $w(x,c)^\dag$. Previous authors have given consideration to the differences between fermionic and bosonic QWs \cite{bib:Matthews11,Sansoni12}. While such differences are valuable to understand, in this paper we will focus on bosonic QWs, given their applicability to present-day optical implementations. Thus, $w(x,c)^\dag$ can be directly interpreted as a bosonic creation operator. The same formalism could be applied to fermionic walkers, but with different commutation relations.

\begin{figure}[!htb]
\includegraphics[width=0.35\columnwidth]{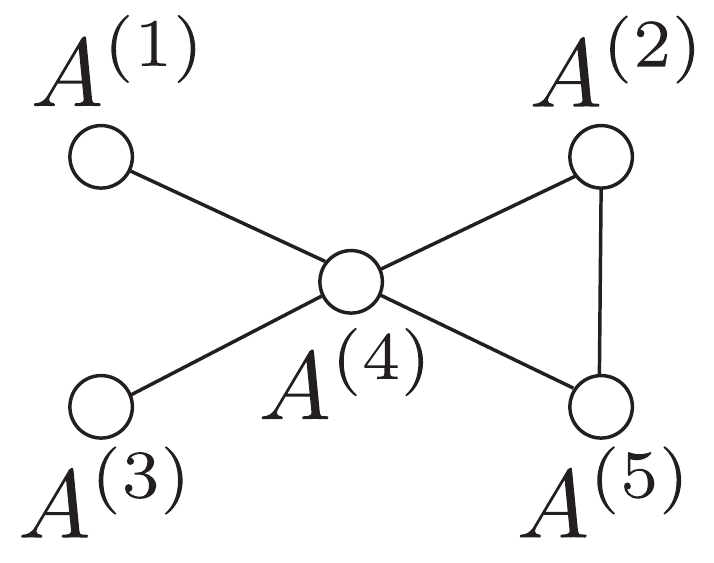}
\caption{Example graph with 5 positions. A unitary coin operator $A^{(i)}$ acts on the neighbourhood of each vertex $i$.} \label{fig:example_graph}
\end{figure}

We define the coin and step operators as,
\begin{eqnarray} \label{eq:coin_step_def}
&C(t):& \, w(x,c)^\dag \mapsto \sum_{j\in n_x} A_{cj}^{(x)}(t) w(x,j)^\dag, \nonumber \\
&S:& \, w(x,j)^\dag \mapsto w(j,x)^\dag.
\end{eqnarray}
Here $n_x$ is the \emph{neighbourhood} of position $x$, i.e. the vertices to which $x$ is connected in the graph, and $A^{(x)}$ is an operator which determines the evolution of the coin value within the respective neighbourhood. An example is illustrated in Fig. \ref{fig:example_graph}. Note that $w(x,c)^\dag\ket{0}$ forms a basis for all single-walker states. Here $\ket{0}$ represents an empty graph, where no walkers are present. In an optical context this corresponds to the vacuum state. The evolution proceeds as $U=\prod_t S\cdot C(t)$.

See also Ref. \cite{bib:Mayer11} for a discussion on multi-particle quantum walks.

%
%

\section{Virtual walkers}

With $n$ walkers in the system the basis states will be of the form $w(x_1,c_1)^\dag w(x_2,c_2)^\dag\dots w(x_n,c_n)^\dag\ket{0}$. We consequently define \emph{virtual walker operators} of the form,
\begin{equation}
v(\vec{x},\vec{c})^\dag \equiv w(x_1,c_1)^\dag w(x_2,c_2)^\dag\dots w(x_n,c_n)^\dag
\end{equation}
(up to normalisation) where $\vec{x}=\{x_1,x_2,\dots,x_n\}$ and $\vec{c}=\{c_1,c_2,\dots,c_n\}$. We refer to these as \emph{virtual walkers} since each set of walker creation operators, while representing multiple particles, can be thought of as a single excitation, inheriting their properties from the underlying bosonic creation operators -- they are indistinguishable, commutative and obey $\bra{0} v(\vec{x},\vec{c}) v(\vec{y},\vec{d})^\dag \ket{0} = \delta_{\vec{x},\vec{y}} \delta_{\vec{c},\vec{d}}$.

\begin{figure}[!htb]
\includegraphics[width=0.7\columnwidth]{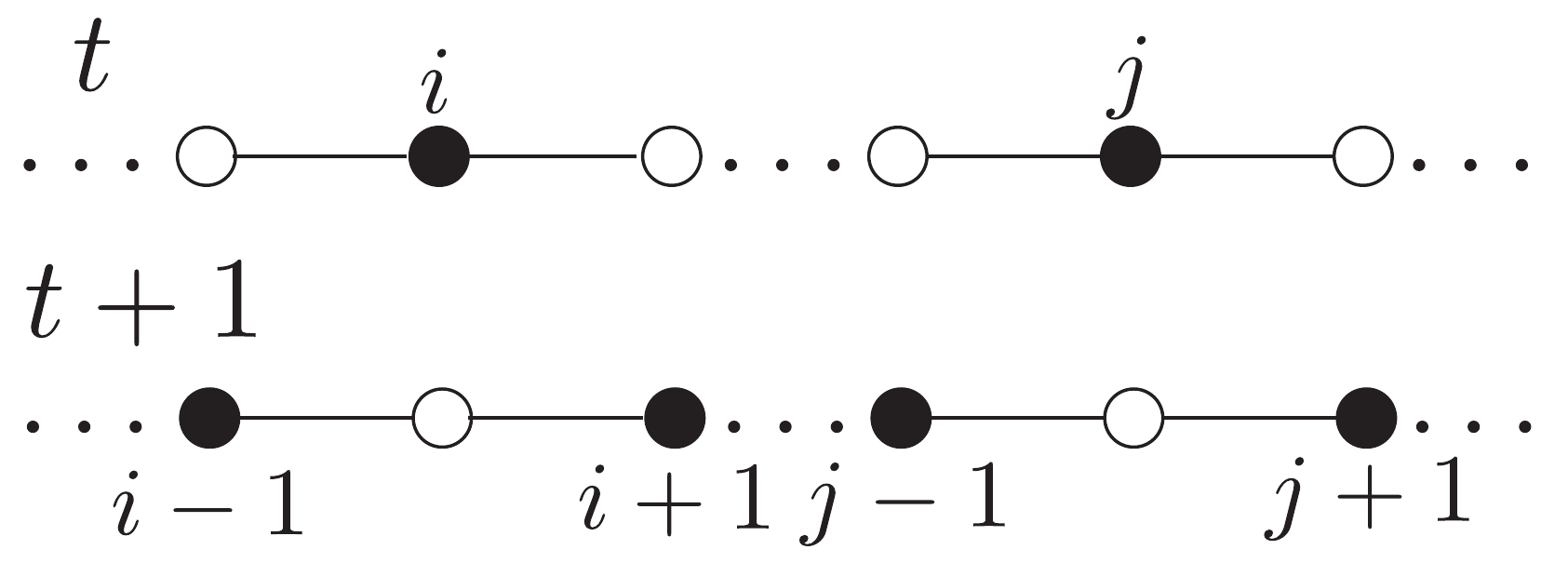}
\caption{Evolution of two walkers at time $t$ to time $t+1$. Black positions are potentially occupied, while white positions are unoccupied. We assume the two walkers are in positions $(i,j)$. Then the allowed transitions can be immediately read off the graph.} \label{fig:cell}
\end{figure}

Let us consider a linear graph as shown in Fig. \ref{fig:cell}(top). With two walkers in the system at positions \mbox{$(i,j)$}, we can immediately read off which transitions between virtual walkers are allowed, Fig. \ref{fig:cell}(bottom). With two walkers it can be seen that the pair of walkers at vertices \mbox{$(i,j)$} may transition to \mbox{$(i+1,j+1)$}, \mbox{$(i+1,j-1)$}, \mbox{$(i-1,j+1)$} and \mbox{$(i-1,j-1)$}. For a linear graph with identical coins at all positions, all $A^{(x)}$ are equal. However, in general they may be distinct, allowing arbitrary graphs structures to be defined with coins of different dimension.

\begin{figure}[!htb]
\includegraphics[width=0.7\columnwidth]{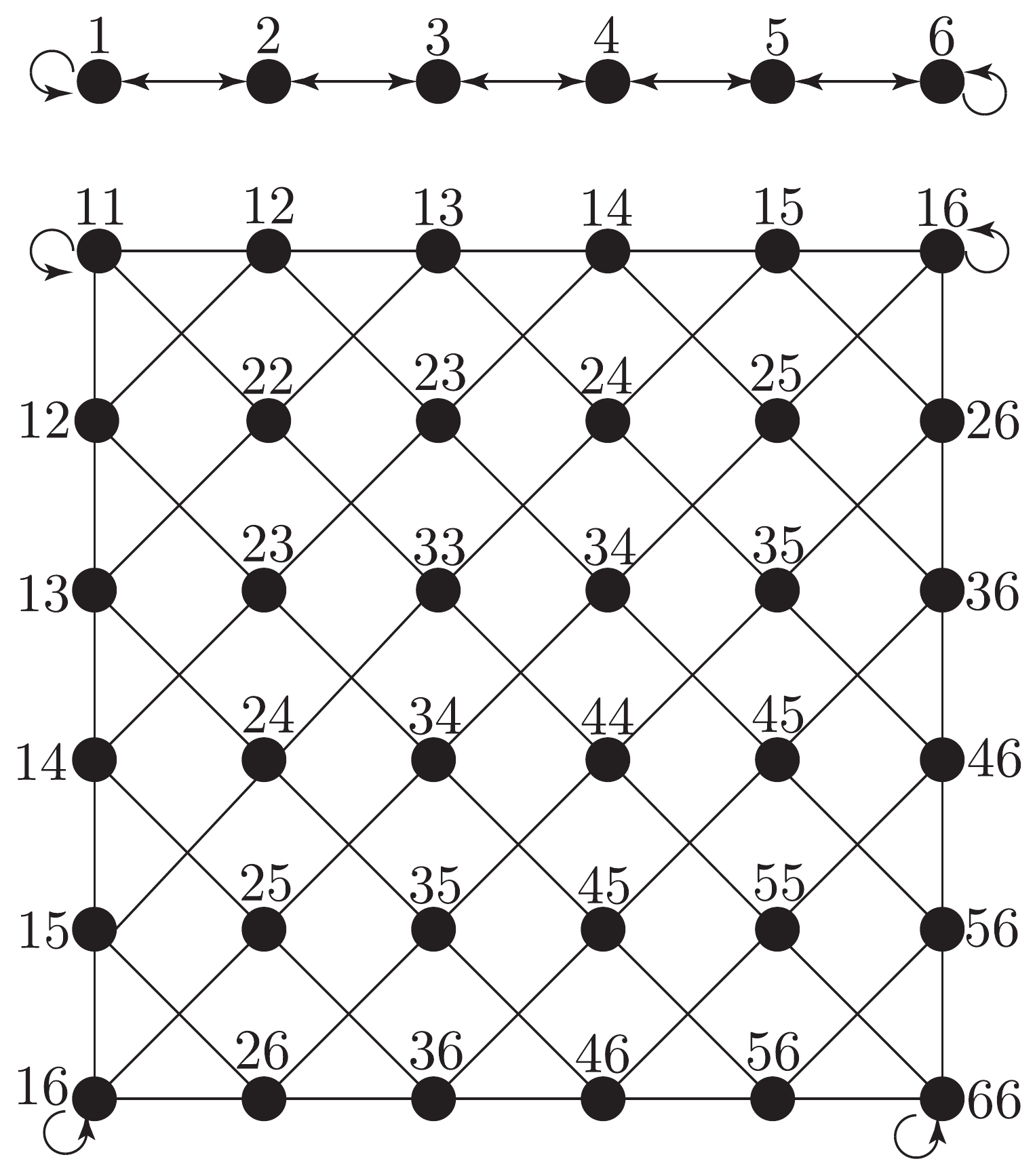}
\caption{(top) A linear walk with 6 vertices. Numbers denote different positions. (bottom) The equivalent walk where two walkers are mapped to a single virtual walker. Numbers represent pairs of positions. Note that the graph is symmetric about the diagonal axis. This is because the particles are indistinguishable, therefore \mbox{$(x_1,x_2)=(x_2,x_1)$}, and all non-diagonal vertices are 2-fold redundant.} \label{fig:6_linear}
\end{figure}

From the allowed transitions we can construct a \emph{virtual graph}. In the example of an underlying linear graph, Fig. \ref{fig:6_linear}(top), it can be seen that the virtual graph is a two-dimensional lattice graph, Fig. \ref{fig:6_linear}(bottom). Thus, the virtual graph, with a single walker, exhibits the same allowed transitions as the underlying graph with two walkers. Generally, in the virtual graph there will be an edge $(i,j)\to (k,l)$ if in the underlying graph there are edges $i\to k$ and $j\to l$. Representing the virtual walkers in a two-dimensional space we can see that the graph is composed of overlapping four pointed stars as the basic building block. Thus, with the addition of a second walker, a linear walk consisting of one walker is transformed into a two-dimensional lattice graph consisting of a single virtual walker. With the addition of a third walker the basic building block will be a three-dimensional star within a three-dimensional lattice, and for $n$ walkers the virtual graph will be an $n$-dimensional lattice. As the number of walkers increases, clearly the size of the virtual graph increases exponentially, even though the underlying linear graph is not increasing in size. Thus, the virtual graph allows us to trade quantum complexity for exponentially growing classical complexity. We have focussed on linear graphs as a simple example, since most present-day optical implementations have been restricted to linear graphs. The described walker operator formalism may be applied to studying arbitrary graph structures. More generically, a connected graph $K_{|G|}$ on graph $G$ with $n$ walkers, will transform into the connected graph $K_{|G|^n}$. The degree of vertex $p_{x_1,x_2,\dots,x_n}$, or equivalently the size of the coin space, in the virtual graph is related to the degree of the vertices $q_y$ in the underlying multi-walker graph by \mbox{$|p_{x_1,x_2,\dots,x_n}| = \prod_{i=1}^{n}|q_{x_i}|$}, which in general grows exponentially with $n$.

\begin{figure*}[!htb]
\includegraphics[width=0.3\textwidth]{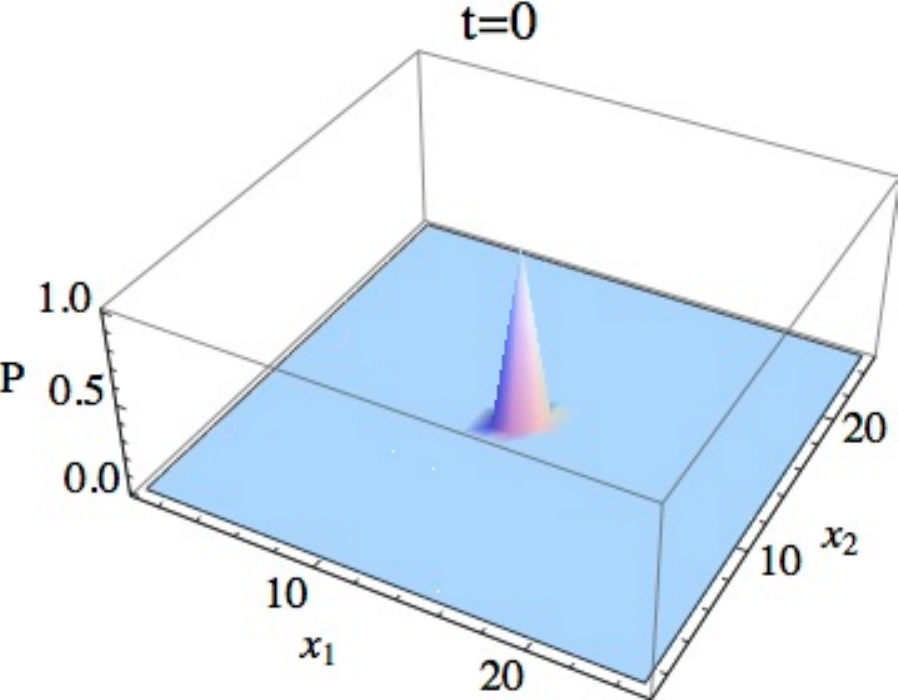} \includegraphics[width=0.3\textwidth]{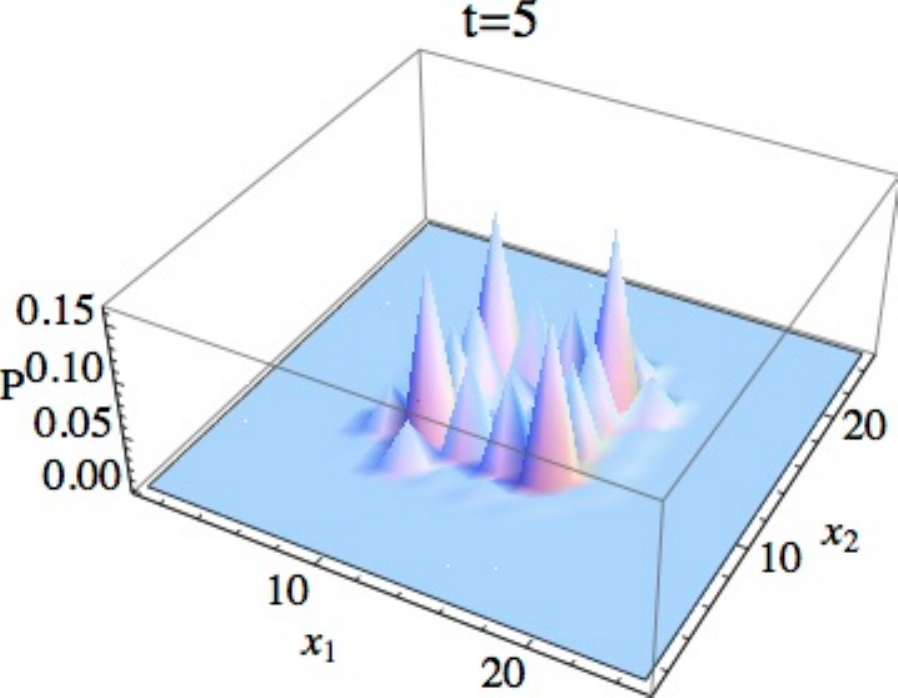} \includegraphics[width=0.3\textwidth]{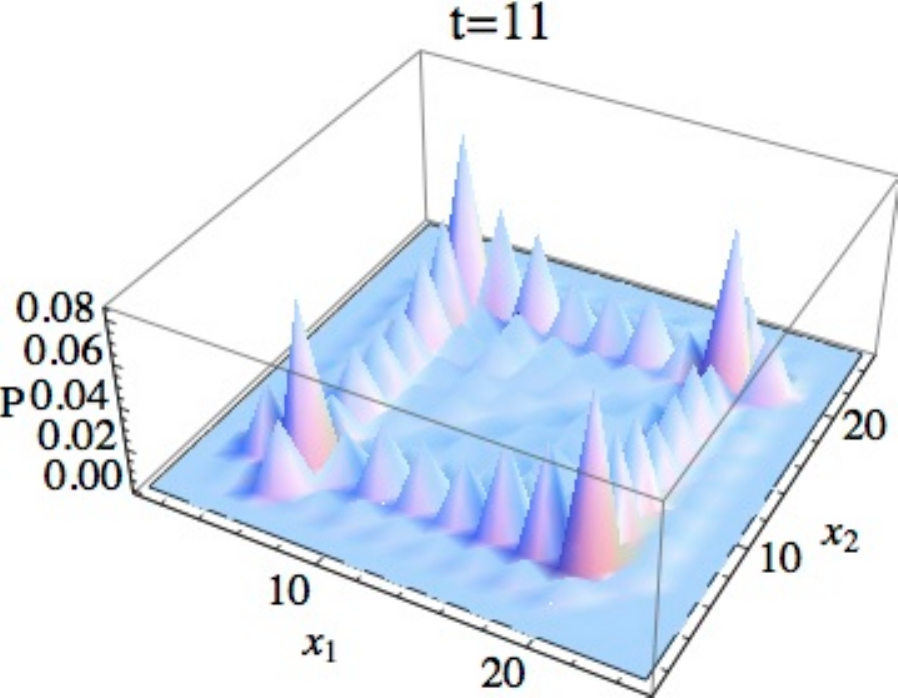}
\caption{(Color online) Probability plots for one walker in a lattice of two spatial dimensions, or equivalently, the coincidences for two walkers on a line (a virtual walker). Because the line is translation invariant, all coins $A^{(x)}$ are equal and are defined as Hadamard operations. Note the walkers/virtual walker exhibit ballistic spreading, a unique feature of quantum mechanical walks.} \label{fig:plots_3d}
\end{figure*}

In Fig. \ref{fig:plots_3d} we show the walker probability plots for two equivalent situations: (1) $P(x_1,x_2)$ is the position of a single walker on a 2D lattice of the form shown in Fig. \ref{fig:6_linear}, where $x_1$ and $x_2$ represent the two spatial dimensions; and, (2) $P(x_1,x_2)$ is the coincidence probability of two walkers on a linear graph, where $x_1$ and $x_2$ denote two distinct positions on a line. Note that because of the isomorphism, virtual walkers on a line exhibit the same properties as a single walker on a lattice. In the graphs in Fig. \ref{fig:plots_3d}, ballistic spreading is observed, an archetypal property of quantum walks \cite{bib:Kempe08}.

The idea of mapping an efficiently sized graph to an exponentially large one was very recently described by Underwood \& Feder \cite{bib:Underwood12}, in which they showed that such such systems can be made universal for quantum computation using Bose-Hubbard type interactions between interacting bosons. We will later demonstrate a similar result for the universality of our virtual graphs via the introduction of defects into the virtual graph.

%
%

\section{Introducing lattice defects}

We now consider some interesting properties of this mapping. First, note that in an arbitrary linear optics network a single photon input state will give the same single click-statistics as weak coherent light. Thus, while two walkers on a line must be implemented using two photons, one walker on a lattice can be implemented with either a single photon or with coherent light. Note that the latter system is purely classical and the former is, in general, highly entangled. Thus, a 2D spatial lattice represents a system, which, while containing no entanglement, is able to simulate a system with entanglement. Importantly however, even though no \emph{entanglement} is present in the latter system, \emph{coherence} is still required.

\begin{figure}[!htb]
\includegraphics[width=0.7\columnwidth]{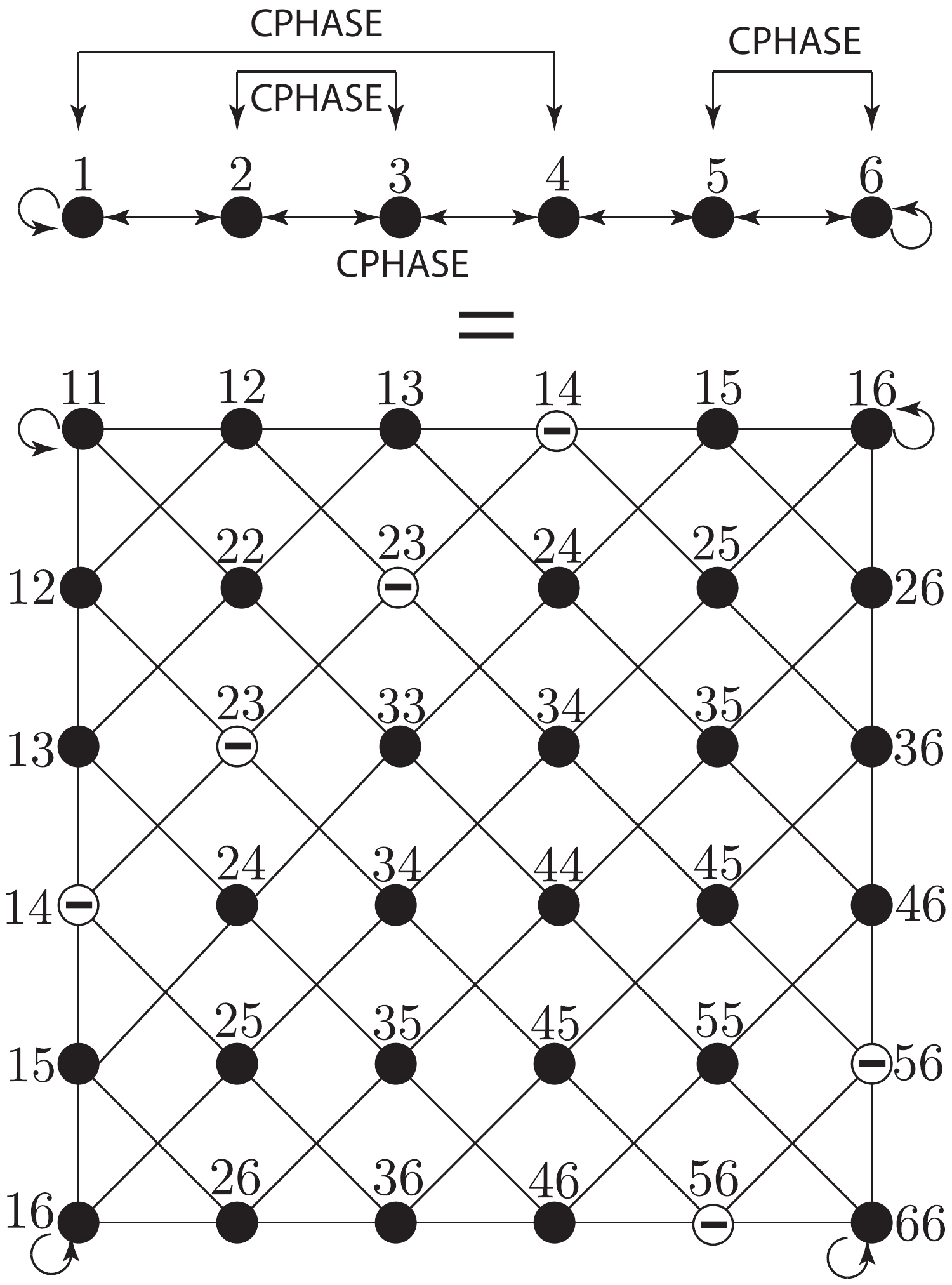}
\caption{An example mapping between entangling {\sc CPHASE} gates on a linear graph with two walkers and non-entangling {\sc PHASE} gates (denoted by `-') in the corresponding virtual graph. An appropriate choice of {\sc CPHASE} gates allows us to efficiently `etch' arbitrary phase-defects into the corresponding virtual graph, subject to the constraint that the virtual graph is symmetric about the diagonal axis. This allows us to generate a larger class of virtual graphs that are in general not translation-symmetric and which may exhibit less trivial behaviour.} \label{fig:defects}
\end{figure}

Next one might ask the question whether entangling \emph{operations} can be implemented in such a system. Suppose we wish to implement a generalisation of the {\sc CPHASE} gate, which applies a phase-flip at position $\{x_1,\dots,x_n\}$ in the virtual graph only if there is a walker at all positions $x_1,\dots,x_n$ in the underlying $n$-walker graph. For example, a two-particle CPHASE gate would be defined as $\mathrm{CPHASE}_{x_1,c_1,x_2,c_2}:\, w(x_1,c_1)^\dag w(x_2,c_2)^\dag \to -w(x_1,c_1)^\dag w(x_2,c_2)^\dag$. In the linear graph with two walkers, this clearly requires an entangling operation -- in fact, in an optical context, it would require a massive non-linear interaction, putting a $\pi$ phase-shift on two-walker terms. However, in the 2D lattice with a single walker this is no longer the case. Rather, we can pick out an individual term of interest and apply a non-entangling {\sc PHASE} gate at the respective position in the virtual graph (see Fig. \ref{fig:defects}). We refer to applications of such localised {\sc PHASE} gates as \emph{lattice defects}. Such defects present us with several interesting properties:

1) We can simulate entanglement with classical states, by accepting an increase in the graph dimensionality. This is interesting from a quantum simulation perspective. Specifically, by accepting an exponential overhead (against the number of walkers) in graph size, we can simulate a multi-walker QW with entangling operations, such as {\sc CPHASE} gates and non-linear interactions, even though no entanglement is present.

2) We can use {\sc CPHASE} gates on a graph with multiple walkers to efficiently `etch' an arbitrary pattern of phase-defects into the corresponding virtual graph. This is particularly interesting as it allows us to break the translation symmetry otherwise present in the virtual graph, opening the possibility to explore more elaborate, non-trivial graph structures with potentially more interesting dynamics. The virtual  graph will always be symmetric about the diagonal axis, since with indistinguishable walkers $\{x_1,x_2\} = \{x_2,x_1\}$. However, modulo this constraint, arbitrary structures of phase-defects can be efficiently implemented with an appropriate choice of {\sc CPHASE} gates.

3) Vertices on the main diagonal of the virtual graph play a special role, since these correspond to both walkers being in the same position. Thus, a phase shift applied to these vertices represents a non-linear interaction where the applied phase is a function of how many walkers reside at the vertex.

\begin{figure*}[!htb]
\includegraphics[width=2\columnwidth]{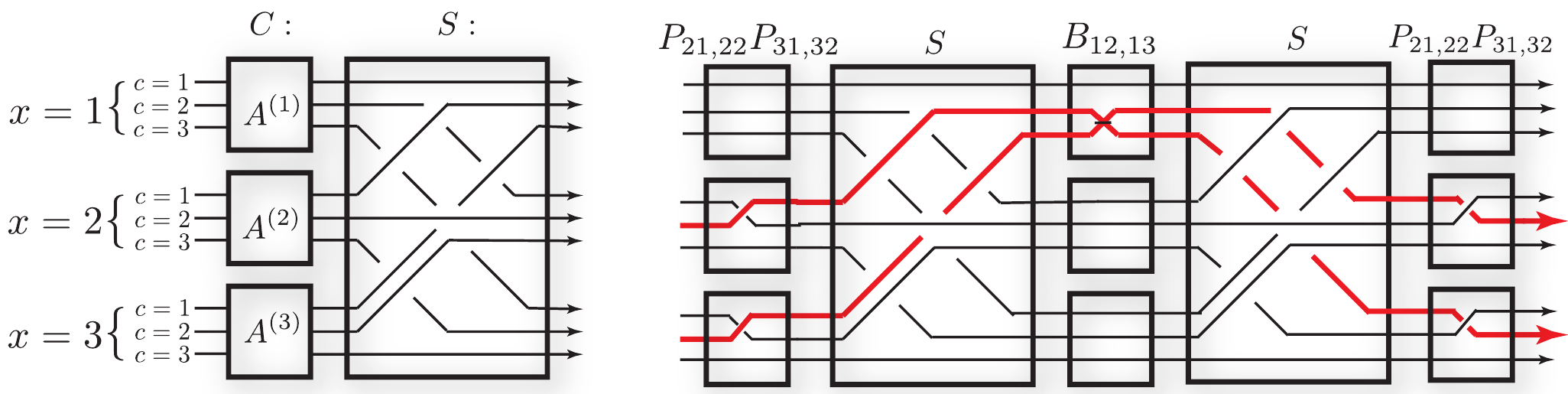}
\caption{(Color online). (left) Mapping an $N=3$ quantum walk to a beamsplitter network. (right) Mapping an arbitrary single beamsplitter operation between modes $(2,2)$ and $(3,2)$, of 9 modes, to an $N=3$ quantum walk. The routing of the modes to which the beamsplitter operation is applied are shown in bold red. First, a choice of permutation coin operators maps the two modes to have the same coin value, after which the step operator maps them to the same position. Then an additional coin applies an arbitrary beamsplitter operation to the two modes. A subsequent step and coin route the two modes back to their original position/coin values.} \label{fig:mapping}
\end{figure*}

4) With the addition of generalised {\sc CPHASE} gates the model becomes universal for QC. To see this we demonstrate that the model can be directly mapped to a qubit system with a universal gate set. Let us begin with a linear walk. Then at each vertex there are two allowed coin states (left and right), $w(x,x+1)^\dag$ and $w(x,x-1)^\dag$. Thus, with a single walker at each vertex in a superposition of the two allowed coin states, we define a set of qubits -- one at each vertex, $\alpha \cdot w(x,x+1)^\dag + \beta \cdot w(x,x-1)^\dag$. Arbitrary single qubit operations can trivially be performed with an appropriate map within the subsequent coin operations ($C$). The final step is to introduce the {\sc CPHASE} gate, the last remaining required gate to form a universal gate set. To implement this gate we begin with a `routing' argument. As an example, Fig. \ref{fig:mapping}(right) illustrates how to map two arbitrary wires to lie within a single coin operator $C$. Now, we allow this $C$ to be more than just a unitary map on the creation operators, but a {\sc CPHASE} operator acting on the appropriate rails within the `bundle' of modes corresponding to a given position (Fig. \ref{fig:mapping}(left)). Note that implementing {\sc CPHASE} gates directly is in general non-deterministic when implemented using just linear optics \cite{bib:KLM01}. Thus, this final step remains the challenging one.

Interestingly, this implies that, when looked at from the perspective of the corresponding virtual graph, an exponentially large regular lattice graph is universal for QC, with an appropriate choice of phase-defects. Without phase-defects the model is unlikely universal for QC since it does not allow for the construction of arbitrary graphs and therefore cannot be mapped to a protocol such as that by Lovett \emph{et al} \cite{bib:Lovett10}. This result on the other hand suggests that full control over the graph is not necessary. Rather, a regular lattice is sufficient, with full control over phase-defects.

In the same way that the QW formalism provided a new approach to visualising and designing quantum algorithms in a graph theoretic context, this alternate approach, based on regular lattices with phase-defects, may provide insight into new approaches to algorithm design.

In Ref. \cite{bib:Schreiber12} Schreiber \emph{et al.} make use of the aforementioned isomorphism to simulate two walkers on a line. They begin by experimentally demonstrating an optical 2D QW, and with the addition of an electro-optic modulator (EOM) are able to apply arbitrary phase defects into the 2D lattice. They subsequently use the isomorphism to simulate two walkers, and are able to simulate both entangling and non-linear interactions, providing a rich test-bed for the simulation of multi-particle dynamics. With appropriate choices of coins and defects they are able to simulate highly entangled two-particle dynamics using just classical light.

%
%

\section{Implications from complexity theory}

We now relate the described multi-walker formalism to two protocols and comment on the implications for complexity theory. Having discussed the potential for multi-walker QWs to achieve exponential complexity, the question is whether this isomorphism can be utilised to perform universal QC, or at least some interesting problems which are intractable on a classical computer. We relate the QW formalism to the protocols by Knill, Laflamme \& Milburn \cite{bib:KLM01}, and Aaronson \& Arkhipov \cite{bib:AaronsonArkhipov10}. With the addition of feedforward the multi-walker QW formalism can be made universal for QC, and in the absence of feedforward is equivalent to {\sc BosonSampling}.

%
%

\subsection{Universal quantum computation using quantum walks?}

Childs \cite{bib:Childs09} first demonstrated that a continuous time QW is universal for QC, and a similar proof exists in the discrete time case \cite{bib:Lovett10}. Here an exponentially sized graph with $O(2^n)$ vertices is required to simulate a qubit circuit comprising $n$ qubits. Of course, in a physical system where position states literally correspond to a physical position, an exponential number of vertices is clearly not possible \cite{bib:RohdeSav10}. However, we have demonstrated that (even without phase-defects) with the addition of multiple walkers the complexity of the system grows exponentially, in terms of the number of available basis states (see Ref. \cite{bib:AaronsonArkhipov10} for further discussion on the complexity of linear optics systems with multiple photons), raising the question as to whether this exponential complexity can be mapped to the Childs \cite{bib:Childs09} or Lovett \emph{et al.} \cite{bib:Lovett10} approach and is universal for QC.

In the work by Childs \cite{bib:Childs09} and Lovett \emph{et al.} \cite{bib:Lovett10}, arbitrary graph structures are required for universal QC. There, \emph{widgets} implementing elementary gate operations are attached to $O(2^n)$ \emph{quantum wires}, which propagate basis states. For example, there is a widget which implements a {\sc CNOT} gate, and widgets for implementing the remaining single qubit operations necessary for universal QC.

Because the number of vertices in the virtual graph is exponentially larger than the underlying graph, and we only have direct control over the underlying graph, it is clear that not \emph{all} combinations of virtual graphs may be constructed. We do not have sufficient control over the isomorphism to generate \emph{arbitrary} virtual graphs. Additionally, to fully utilise an exponentially large virtual graph one must have control over an exponential number of coin parameters. Since we can only directly implement coins on the underlying graph it is clear that not all combinations of coins in the virtual graph can be implemented. We therefore suggest that, without phase-defects, universal QC is unlikely using the multi-walker formalism since it cannot be mapped to arbitrary exponentially large single-walker walks. Indeed, a simple counting argument gives stronger weight to the claim that multi-walker QWs cannot replicate a Lovett \emph{et al.} \cite{bib:Lovett10} protocol. Suppose the underlying graph contains $k$ vertices. Then, there are $k$ locations at which coins can be applied. Similarly, in the corresponding virtual graph there are also $k$ allowed unique combinations of locations where coins may be applied. In the Lovett \emph{et al.} \cite{bib:Lovett10} approach, the walk is defined by an exponential number of locations at which coins may be applied. Thus, if all exponentially sized graphs are to be simulated then, if they can be simulated at all, an exponential number of time steps will be required simply to allow for the desired combinations of coin operators, suggesting that \emph{efficient} universal QC is not possible. If efficient universal QC is possible on a virtual lattice \emph{without} phase defects, this would imply that a quantum walk with $N$ basis states is universal with control over just $O(\mathrm{log}\,N)$ coins.

Note that the above discussion is not a formal proof that a multi-walker quantum walk is not universal for quantum computing, but rather an intuitive plausibility argument based on a counting approach.

%
%

\subsection{Relation to linear optics quantum computing}

Although unlikely to be universal for QC (e.g. as proposed by Knill, Laflamme \& Milburn (KLM) \cite{bib:KLM01}), multi-walker QWs can efficiently implement the {\sc BosonSampling} algorithm proposed by Aaronson \& Arkhipov (AA) \cite{bib:AaronsonArkhipov10}. We first show that multi-walker QWs \emph{can} efficiently simulate arbitrary linear optics networks, and then relate this to KLM and AA.

%
%

\subsubsection{Mapping quantum walks to optical networks}

From Eq. \ref{eq:coin_step_def} it is easy to write down a mapping of a quantum walk to a beamsplitter network. Note that the coin operator acts individually on each `bundle', where a bundle of modes represents a walker position. Reck \emph{et al.} \cite{bib:Reck94} showed that an arbitrary unitary map of the form $a_i^\dag \to \sum_j U_{ij} a_j^\dag$ has an efficient decomposition into beamsplitters and phase-shifters. Thus the operation $A^{(x)}$ acting on each bundle can be efficiently constructed. Then it can easily be seen that Fig. \ref{fig:mapping}(left) is an efficient decomposition of an arbitrary quantum walk into a beamsplitter and phase-shifter network.

%
%

\subsubsection{Mapping arbitrary optical networks to a quantum walk}

Mapping in the opposite direction, from a beamsplitter and phase-shifter network to a quantum walk is not as trivial. We aim to show that a QW can act as a decomposition for an arbitrary linear optics network. It suffices to show that an efficient number of coin and step operators can be used to apply beamsplitter/phase-shifter operations between arbitrary pairs of modes. To do this we first use permutations within the first coin operator. These permutations, followed by the step operator, route two arbitrary modes to the same position `bundle'. Once both modes have been mapped to the same position an arbitrary beamsplitter/phase-shifter operation can be applied to the corresponding two modes with an appropriate choice of $A^{(i)}$. Then, another application of the same step and permutation operators route the respective modes back to their original coin and position states. In general, multiple operations can be performed in parallel. Formally, the decomposition can be expressed,
\begin{eqnarray}
B_{x_1c_1,x_2c_2} &=& P_{x_11,x_1c_1}\cdot P_{x_21,x_2c_2}  \cdot S \cdot \nonumber \\
&& B_{1x_1,1x_2} \cdot S \cdot P_{x_11,x_1c_1} \cdot P_{x_21,x_2c_2},
\end{eqnarray}
where $P_{ij,kl}$ is a permutation between modes $(i,j)$ and $(k,l)$, $S$ is the usual step operator, and $B_{ij,kl}$ is the beamsplitter operation between modes $(i,j)$ and $(k,l)$. Here $P$ and $B$ are the choices of coin operators, so the sequence of operators still consists of iterations of coin and step operators. An example is illustrated in Fig. \ref{fig:mapping}(right). Finally, we know from Reck \emph{et al.} \cite{bib:Reck94} that a polynomial number of iterations of coin and step operators are now sufficient to decompose an arbitrary linear optics network into a QW. Thus, a QW is an efficient decomposition for arbitrary unitary maps.

%
%

\subsubsection{Relation to KLM and AA}

We now briefly discuss the relationship between multi-walker QWs and LOQC. We discuss two different approaches to LOQC -- KLM and AA. These two models are subtly different. The KLM approach requires arbitrary linear optics networks and the addition of multi-photon input states, measurement, post-selection and fast feed-forward. This approach is known to be universal for QC (i.e. it can solve \textbf{BQP}-complete problems). The AA approach on the other hand also requires general linear optics networks and multi-photon input states, but mitigates the necessity for post-selection and feed-forward. This approach implements the so-called {\sc BosonSampling} problem, which is strongly believed to be classically hard (residing in the class \textbf{BosonSampP}-complete) \cite{bib:AaronsonArkhipov10}. While far simpler and significantly experimentally less demanding, this approach is neither known nor believed to be universal for QC \cite{bib:AaronsonArkhipov10}. Note that both approaches require arbitrary beamsplitter networks. We have shown in the previous section that QWs act as a decomposition for arbitrary linear optics networks. Thus, it follows that multi-walker QWs can be used to implement both KLM and AA.

In the case of KLM the QW must be complemented by measurement and feed-forward. Thus the corresponding QW consists not only of coin and step operators, but also measurement operators, and the evolution of the system can be represented as $\prod_t M(t) \cdot S \cdot C(t,m_{t-1})$, where $M$ is a measurement operator, and $m_t$ denotes the measurement outcome at time $t$.

AA on the other hand can be implemented using a QW in the more conventional way of iterative applications of just coin and step operators. We have demonstrated that in the absence of feedforward, multi-walker QWs are unlikely to be universal for QC, since there is no general mapping to the QW protocol of Lovett \emph{et al} \cite{bib:Lovett10}. This gives further weight to the claim that {\sc BosonSampling} is likely not universal for QC. As discussed earlier, if {\sc BosonSampling} were to be universal for QC, this would imply that the protocol of Lovett \emph{et al.} \cite{bib:Lovett10} is universal for QC even when acting on exponentially sized regular lattice structures of size $N$ with control over only $O(\mathrm{log}\, N)$ coins. We can establish the condition that {\sc BosonSampling} is universal for efficient QC only if a single-walker walk on exponentially large graphs can be implemented using time-dependent periodic graph structures, with an efficient number of time steps. Conversely, because of the isomorphism between multi-walker QWs and arbitrary linear optics networks, the work of AA provides a useful application for multi-walkers QWs. Although such regular lattice structures are unlikely universal for QC, the question is raised whether nonetheless \emph{some} interesting problems can be solved on such structures.

%
%

\section{Entangled input states}

In conventional approaches to QC, such as the circuit and cluster state \cite{bib:Raussendorf01,bib:Raussendorf03} models, it is known that initially entangled states act as a useful resource to the benefit of QC. Also they provide an advantage in QW algorithms, as for the example the graph isomorphism problem \cite{bib:Berry11a}. One might ask the question whether the same applies in the context of our multi-walker QW model. Specifically, we know that with the addition of {\sc CPHASE} gates our model becomes universal. However, such entangling gates are both challenging and non-deterministic in the optical context. Thus, the question is whether the non-trivial gates can be commuted to the beginning of the circuit such that the challenging part of the problem is transferred to a resource state preparation problem after which the algorithm is `easily' implemented using just the passive, and deterministic, sequence of coin and step operators. Here we show that this is not the case. We present a simple example of a circuit which illustrates that, in general, such resource state preparation is as difficult as implementing the {\sc CPHASE} gates within the circuit. Thus, the idea of commuting {\sc CPHASE} gates to the beginning of the circuit and treating it as a resource state preparation problem is of little practical benefit.

\begin{figure}[!htb]
\includegraphics[width=\columnwidth]{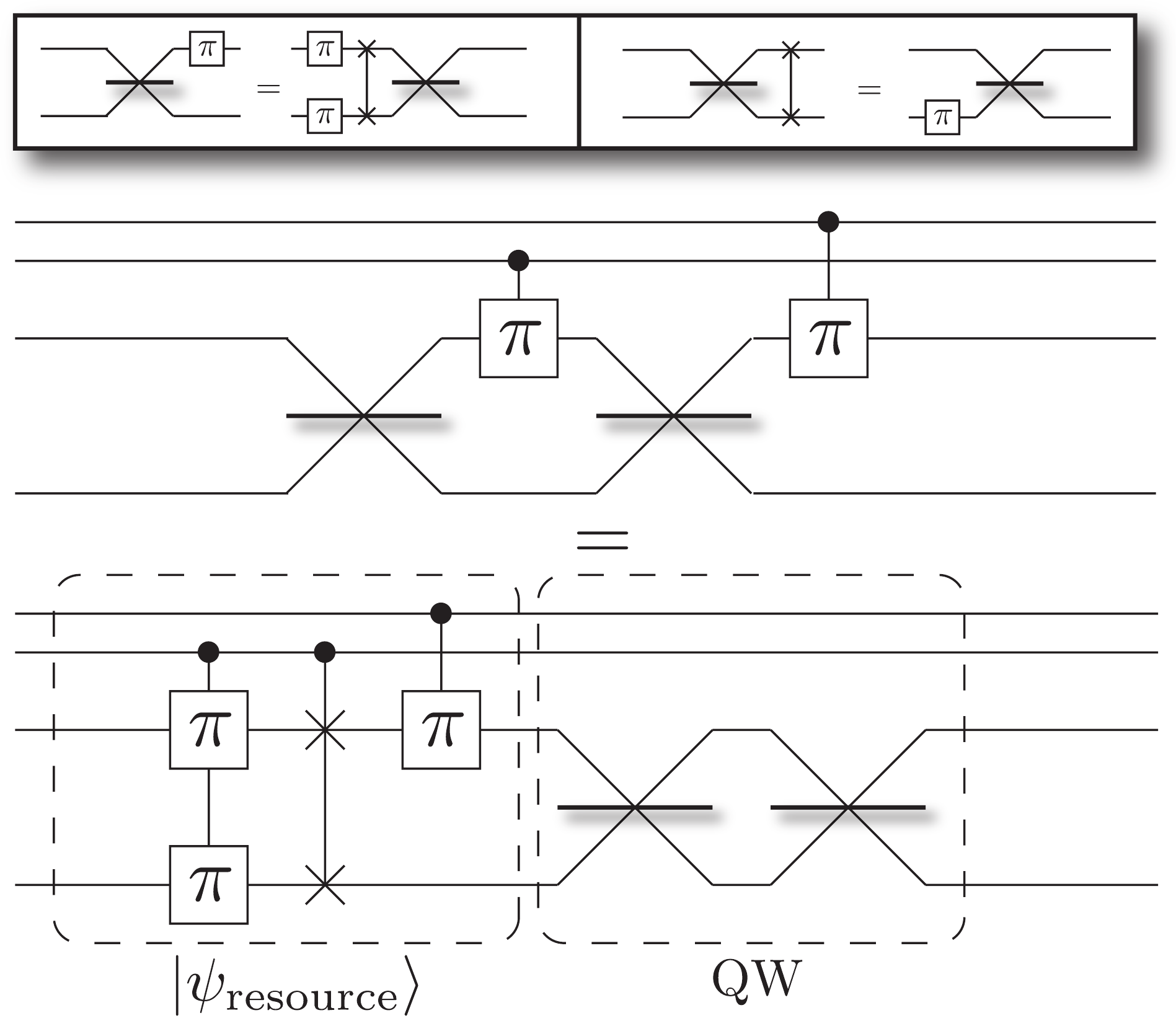}
\caption{(inset top) Commutation relation for a $\pi$-phase gate and a beamsplitter, and a swap gate and a beamsplitter. (bottom) Commuting controlled-$\pi$ gates within a trivial QW to the beginning of the circuit, such that the circuit consists of an entangled resource state and a QW without any controlled gates.} \label{fig:commute}
\end{figure}

In Fig. \ref{fig:commute} we present an extremely simple circuit consisting of two control wires and two additional wires to which two consecutive beamsplitters are applied. After each beamsplitter comes a {\sc CPHASE} gate, controlled by one of the two control wires. Using the commutation relations in Fig. \ref{fig:commute}(inset top) we commute the {\sc CPHASE} gates to the beginning of the circuit and treat all the components preceding the QW as a resource state $\ket{\psi_\mathrm{resource}}$. Evidently, the number of controlled gates in the state preparation procedure is as large as the number of controlled gates we had in the original circuit. On one hand this suggests that resource state preparation does not mitigate the difficulties inherent in implementing controlled gates within the circuit. On the other hand, because all the non-deterministic gates occur at the beginning of the circuit, they can be implemented `offline', a trick which is well known for optical cluster states \cite{bib:Nielsen04}. However, in the optical context, entangling controlled gates are extremely challenging. Thus we suggest that the trick of resource state preparation does not represent an efficient solution to the problem.

\begin{figure}[!htb]
\includegraphics[width=0.5\columnwidth]{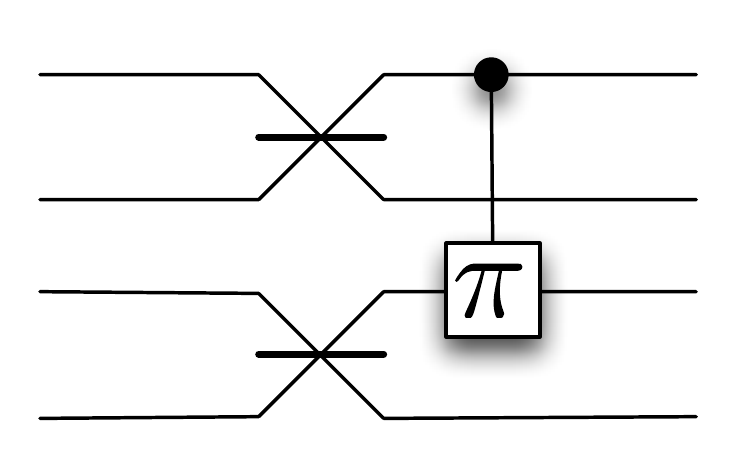}
\caption{A simple example of a beamsplitter network with a single CPHASE gate between two modes. There is no trivial commutation relation to commute the CPHASE gate to before the beamsplitters. Commutation would require a four-mode entangling operation.}
\label{fig:non_commute}
\end{figure}

It should also be noted that while in our simple toy example commuting all controlled gates to the beginning of the circuit is possible, in general this is not the case. It is trivial to draw circuits where the described commutation relations do not allow for all entangling operations to be commuted to the beginning of the circuit. For example, consider the simple circuit shown in Fig. \ref{fig:non_commute}. Here no trivial commutation relation exists enabling the CPHASE gate to be commuted to before the beamsplitters. Commutation would involve a four-mode entangling operation. Nonetheless, it may be the case that there is a subset of interesting problems for which commutation is possible and the resource state preparation is more practical than implementing controlled gates within the circuit.

%
%

\section{Conclusion}

We have shown that adding walkers to a quantum walk system is equivalent to a single walker on an exponentially growing virtual lattice. Although the complexity of the system grows exponentially, it is unlikely to be universal for quantum computation as the resulting graph has inherent symmetries that limit our ability to engineer arbitrary graph structures, unless CPHASE gates are introduced, which generate arbitrary lattice defects. However, in the absence of CPHASE gates, a multi-walker quantum walk is isomorphic to the {\sc BosonSampling} problem, which is of practical interest, and within reach of near-term experiments. Our arguments give further weight to the assertion that {\sc BosonSampling} is not universal for quantum computation, since there is no equivalence to the scheme for universal quantum computation using quantum walks. Nonetheless, multi-walker QWs are likely capable of tasks classically intractable and therefore of practical interest. We demonstrated that single-walker 2D walks can simulate two-walker 1D walks and therefore a strictly non-entangling system (i.e. one walker simulated with weak coherent light) can simulate a highly entangling system. This observation paves the way for elementary demonstrations of quantum walks with multiple walkers, via the addition of extra dimensions. We also demonstrated that with the addition of entangling gates, such as {\sc CPHASE} gates, the symmetry of the corresponding virtual graph can be broken by adding defects to the lattice, paving the way for non-trivial graph structures to be studied. Of particular interest, with the addition of {\sc CPHASE} gates, or equivalently phase-defects in the corresponding virtual graph, the system becomes universal for quantum computation, despite the fact that we do not have full control over the graph structure. Rather, a regular lattice combined with full control over phase-defects is sufficient for universal quantum computation.

%
%

\begin{acknowledgments}
We thank Timothy Ralph, Scott Aaronson and Aur{\' e}l G{\' a}bris for helpful discussions. This research was conducted by the Australian Research Council Centre of Excellence for Engineered Quantum Systems (Project number CE110001013). We acknowledge support by the grant MSM 6840770039 of the Czech Ministry of Education and the Doppler Institute, FNSPE CTU in Prague.
\end{acknowledgments}

%
%

\bibliography{paper}

\end{document}